\documentclass[%
 reprint,
 showpacs,
 showkeys,
 preprintnumbers,
 amsmath,amssymb,
 aps,
 pra,
  longbibliography,
 ]{revtex4-1}

\usepackage[breaklinks=true,colorlinks=true,anchorcolor=blue,citecolor=blue,filecolor=blue,menucolor=blue,pagecolor=blue,urlcolor=blue,linkcolor=blue]{hyperref}
\usepackage{graphicx}
\usepackage{xcolor}
\usepackage[left]{eurosym}
\usepackage{url}

\begin{document}

\title{Indeterminism and Randomness Through Physics}
\thanks{Hector Zenil posted the following five questions:
``Why were you initially drawn to the study of computation and randomness?'',
``What have we learned?'',
``What don't we know (yet)?'',
``What are the most important open problems in the field?'',
``What are the prospects for progress?'',
 at URL {\tt http://www.mathrix.org/experimentalAIT/RandomnessBook.htm}, accessed on May 1st, 2009.}

\author{Karl Svozil}
\affiliation{Institute of Theoretical Physics, Vienna
    University of Technology, Wiedner Hauptstra\ss e 8-10/136, A-1040
    Vienna, Austria}
\email{svozil@tuwien.ac.at} \homepage[]{http://tph.tuwien.ac.at/~svozil}

\date{\today}

\begin{abstract}
Despite provable unknowables in recursion theory, indeterminism and randomness in physics is confined to conventions, subjective beliefs and preliminary evidence. The history of the issue is very briefly reviewed, and answers to five questions raised by Hector Zenil are presented.
\end{abstract}

\maketitle

It is not totally unreasonable to speculate if and why  the universe we experience with our senses and brains appears to be ``(un)lawful.''
Indeed, the ``unreasonable effectiveness of mathematics in the natural sciences''~\cite{wigner} seems mind-boggling and tantamount to our (non)existence.
Beyond belief, there do not seem to exist {\it a priori} answers to such questions which would be forced upon us, say maybe by consistency constraints.
But then, why should consistency and logic be considered {\it sacrosanct}?

In view of the opaqueness of the issues, a fierce controversy between supporters and opponents of
a ``clockwork universe'' {\it versus} ``cosmic chaos'' has developed from antiquity onwards ---
cf., e.g., Aristotle's comments on the Pythagoreans in {\em Physics},
as well as Epicurus' {\em Letter to Menoeceus}.
Indeed, for the sake of purported truth, many varieties of conceivable mixtures of determinism and chance have been claimed and repudiated.

The author has argued elsewhere~\cite{2002-cross} that there are many emotional reasons
(not) to believe in a(n) (in)deterministic universe: does it not appear frightening
to be imprisoned by remorseless, relentless predetermination; and, equally frightening,
to accept one's fate as being contingent on total arbitrariness and chance?
What merits and what efforts appear worthy at these extreme positions,
which also unmask freedom, self-determination and human dignity as an idealistic illusion?

In order to disentangle the scientific discussion of topics such as (in)determinism, or realism {\em versus} idealism,
from emotional overtones and possible bias, it might not be totally unreasonable to allow oneself
the contemplative strategy of {\em evenly-suspended attention} outlined by Freud~\cite{Freud-1912}:
Nature is thereby treated as a ``client-patient,'' and whatever comes up is accepted ``as is,'' without any
immediate emphasis or judgment\footnote{
In Ref.~\cite{Freud-1912}, Freud admonishes analysts to be aware of the dangers caused by { ``$\ldots$~temptations to project,
what [[the analyst]] in dull self-perception recognizes as the peculiarities of his own personality,
as generally valid theory into science~$\ldots$''} (In German: {\em ``Er wird leicht in die Versuchung geraten,
was er in dumpfer Selbstwahrnehmung von den Eigent{\"u}mlichkeiten seiner eigenen Person erkennt,
als allgemeing{\"u}ltige Theorie in die Wissenschaft hinauszuprojizieren~$\ldots$~.''})}.

In more recent history, the European Enlightenment (illuminating also wide areas across the oceans)
has brought about the belief of total causality and almost unlimited predictability, control, and manipulative capacities.
Subsequently, the {\em principle of sufficient reason} came under pressure at two independent frontiers:
Poincar\'e's discovery of instabilities in classical many-body motion~\cite{Diacu96},
as already envisioned by Maxwell~\cite[pp.~211-212]{Campbell-1882}, is now considered as a precursor to {\em deterministic chaos},
in which the information ``held'' in the initial value ``unfolds'' through a deterministic process.
Note that, with probability one, an arbitrary real number representing the initial value,
which is  ``grabbed'' (facilitated by the axiom of choice) from the ``continuum urn,'' is
provable random in the sense of algorithmic information theory~\cite{MartinLöf1966602,calude:02,ch6}; i.e.,
in terms of algorithmic incompressibility as well as of the equivalent statistical tests.
Moreover, for entirely different reasons, if one encodes universal computation into a system on $n$ bodies, then by reduction (cf. below) to the halting problem
of recursion theory~\cite{rogers1,davis,Barwise-handbook-logic,enderton72,odi:89,Boolos-07},
certain observables become provable unknowable~\cite{svozil-2007-cestial}.

A second attack against determinism erupted through the development of quantum theory.
Despite fierce resistance of Einstein\footnote{
In a letter to Born, dated December~12th, 1926~\cite[p.~113]{born-69}, Einstein expressed his conviction,
``In any case I am convinced that he [[the Old One]] does not throw dice.''
(In German: {\em ``Jedenfalls bin ich {\"{u}}berzeugt, dass der [[Alte]] nicht w{\"{u}}rfelt.''})},
Schr\"odinger and De Brogli,
Born expressed the new quantum {\it canon},
repeated by the ``mainstream'' ever after~\cite{zeil-05_nature_ofQuantum}, as follows
(cf. Ref.~\cite[p.~866]{born-26-1}, English translation in \cite[p.~54]{wheeler-Zurek:83})\footnote{
{\em ``Vom Standpunkt unserer Quantenmechanik gibt es keine Gr\"o\ss e, die im {\em Einzelfalle} den Effekts eines Sto\ss es
kausal festlegt; aber auch in der Erfahrung haben wir keinen Anhaltspunkt daf\"ur, da\ss~ es innere Eigenschaften
der Atome gibt, die einen bestimmten Sto\ss erfolg bedingen.
Sollen wir hoffen, sp\"ater solche Eigenschaften
[[$\ldots$]] zu entdecken und im Einzelfalle zu bestimmen?
Oder sollen wir glauben, dass die \"Ubereinstimmung von Theorie und Erfahrung
in der Unf\"ahigkeit, Bedingungen f\"ur den kausalen Ablauf anzugeben, eine pr\"astabilisierte Harmonie ist,
die auf der Nichtexistenz solcher Bedingungen beruht?
Ich selber neige dazu,die Determiniertheit in der atomaren Welt aufzugeben.''
}
}:
\begin{quote}
{  ``From the standpoint of our quantum mechanics, there is no quantity
which in any individual case causally fixes the consequence of the collision;
but also experimentally we have so far no reason to believe that there are some inner properties of the atom
which condition a definite outcome for the collision.
Ought we to hope later to discover such properties [[$\ldots$]]  and determine them in individual cases?
Or ought we to  believe that the agreement of theory and experiment --- as to the impossibility
of prescribing conditions for a causal evolution --- is a pre-established harmony founded on the nonexistence of such conditions?
I myself am inclined  to give up determinism in the world of atoms.''
}
\end{quote}
More specifically, Born offers a mixture of (in)determinism: while postulating a probabilistic behavior of individual particles,
he accepts a deterministic evolution of the wave function
(cf. \cite[p.~804]{born-26-2}, English translation in \cite[p.~302]{jammer:89})\footnote{
{\em  ``Die Bewegung der Partikel folgt Wahrscheinlichkeitsgesetzen,
die Wahrscheinlichkeit selbst aber breitet sich im Einklang mit dem Kausalgesetz  aus.
[Das hei\ss t, da\ss~ die Kenntnis des Zustandes in allen Punkten in einem Augenblick
die Verteilung des Zustandes zu allen sp{\"a}teren Zeiten festlegt.]''
}
}:
\begin{quote}
{  ``The motion of particles conforms to the laws of probability, but the probability itself
is propagated in accordance with the law of causality.
[This means that knowledge of a state in all points in a given time determines the distribution of
the state at all later times.]''
}
\end{quote}

In addition to the indeterminism associated with outcomes of the measurements of single quanta,
there appear to be at least two other types of quantum unknowables.
One is complementarity, as first expressed by Pauli~\cite[p.~7]{pauli:58}.
A third type of quantum indeterminism was discovered by studying quantum probabilities,
in particular the consequences of Gleason's theorem~\cite{Gleason}: whereas the classical probabilities
can be constructed by the convex sum of all two-valued measures associated with classical truth tables,
the structure of elementary yes--no  propositions in quantum mechanics associated with projectors in three- or higher-dimensional Hilbert spaces
do not allow any two-valued measure~\cite{specker-60,kochen1}.
One of the consequences thereof is the impossibility of a consistent co-existence of the outcomes of all conceivable quantum observables
(under the {\em noncontextuality} assumption~\cite{bell-66} that measurement outcomes are identical if they ``overlap'').

Parallel to these developments in physics, G\"odel~\cite{godel1} put an end to finitistic speculations in mathematics
about possibilities to encode all mathematical truth in a finite system of rules.
The recursion theoretic, formal unknowables exhibit a novel feature: they present {\em provable} unknowables in the fixed axiomatic system in which they are derived.
(Note that incompleteness and undecidability exist always relative to the particular formal system or model of universal computation.)
From ancient times onwards, individuals and societies have been confronted with a pandemonium of unpredictable behaviors and occurrences in their environments,
sometimes resulting in catastrophes. Often these phenomena were interpreted as ``God's Will.''
In more rationalistic times, one could pretend without presenting a formal proof that certain unpredictable behaviors are in principle deterministic,
although the phenomena cannot be predicted ``for various practical purposes'' (``epistemic indeterminism'').
Now provable unknowables make a difference by being immune to these kinds of speculation.
The halting problem in particular demonstrates the impossibility to predict the behavior of deterministic systems in general;
it also solves the induction (rule inference) problem to the negative.

In order to be able to fully appreciate the impact of recursion theoretic
undecidability on physics~\cite{wolfram84,kanter,moore,wolfram85b,dc-d91a,dc-d91b,suppes-1993,svozil-93,1994IJTP...33.1085H,casti:94-onlimits_book,casti:96-onlimits,barrow-impossibilities,CalCamSvo-Stef-1995,PhysRevE.65.016128},
let us sketch an algorithmic proof of the undecidability of the halting problem; i.e.,
the decision problem of whether or not
a program $p$ on a given finite input finishes running (or will reach a particular halting state) or will run forever.
The proof method will use a {\it reductio ad absurdum}; i.e., we assume the existence of a {\em halting algorithm} $h(p)$ deciding the halting problem of $p$,
as well as some trivial manipulations; thereby deriving a complete contradiction.
The only alternative to inconsistency appears to be the nonexistence of any such halting algorithm.
For the sake of contradiction, consider an agent $q(p)$ accepting as input an arbitrary program (code) $p$.
Suppose further that it is able to consult a halting algorithm $h(p)$, thereby producing the {\em opposite} behavior of $p$:
whenever $p$ halts, $q$ ``steers itself'' into the halting mode;
conversely, whenever $p$ does not halt, $q$ forces itself to halt.
A complete contradiction results from $q$'s behavior on itself, because whenever $q(q)$ detects (through $h(q)$) that it halts,
it is supposed not to halt;
conversely if $q(q)$ detects  that it does not halt, it is supposed to halt.
Finally, since all other steps in this ``diagonal argument'' with the exception of $h$ are trivial,
the contradiction obtained in applying $q$ to its own code proves that any such program --- and
in particular a halting algorithm $h$ --- cannot exist.

In physics, analogous arguments embedding a universal computer into a physical substrate yield provable undecidable observables
{\em via} {\em reduction to the halting problem}.
Note that this argument  neither means that the system does not evolve deterministically on a step-by-step basis,
nor implies that predictions are provable impossible for all cases; that would be clearly misleading and absurd!
A more quantitative picture arises if we study the potential growth of ``complexity'' of deterministic systems
in terms of their maximal capability to ``grow'' before reaching a halting state through the Busy Beaver function~\cite{rado,chaitin-ACM,dewdney,brady}.
Another consequence is the recursive unsolvability of the general induction (or rule inference~\cite{go-67,blum75blum,angluin:83,ad-91,li:92})
problem for deterministic systems.
As an immediate consequence of these findings it follows that no general algorithmic rule or operational method~\cite{bridgman50} exists
which could ``extract'' some rather general law from a (coded) sequence.
(Note again that it still may be possible to extract laws from ``low-complex'' sequences; possibly with some intuition and additional information.)
Nor can there be certainty that some sequence denominated ``random'' is not generated by a decompression algorithm which makes it formally nonrandom;
a fact well known in recursion and algorithmic information theory~\cite{chaitin3,calude:02} but hardly absorbed by the physics community.
Thereby,  to quote {\it Shakespeare's Prospero,} any claims of absolute (``ontological'') randomness decay into ``thin air.''
Of course, one could still vastly restrict the domain of possible laws and {\em define} a source to be random
if it ``performs well'' with respect to the associated, very limited collection of statistical tests,
a strategy adapted by the {\it Swiss Federal Office of Metrology}\footnote{
Cf. the {\it Certificate of Conformity No 151-04255}}.

Despite the formal findings reviewed above, which suggest that claims of absolute indeterminacy cannot be proven but represent subjective beliefs,
their predominance in the physics community can be understood, or rather motivated,
by the obvious inability to account for physical events, such as the outcomes of certain quantum measurements,
e.g., radioactive decays~\cite{Kragh-1997AHESradioact,Kragh-2009_RePoss5}, deterministically.
Why this effective incapacity to predict individual outcomes or time series of measurement data should be different from other ``classical'' statistical sources of randomness
--- even when complementarity and value indefiniteness is taken into account --- remains an open question,
at least from a formal point of view.

For the sake of explicit demonstration, let us consider a particular method of generation of a sequence from single quantum outcomes~\cite{svozil-2006-ran}
by combination of source and beam splitter~\cite{svozil-qct,rarity-94,zeilinger:qct,stefanov-2000,VincentJacques02162007,0256-307X-21-10-027,fiorentino:032334,wang:056107,svozil-2009-howto}.
Ideally (to employ quantum complementarity as well as quantum value indefiniteness), a system allowing three or more outcomes is prepared to be in a particular
pure state ``contained'' in a certain context (maximal observable~\cite{svozil-2008-ql} or block~\cite{greechie:71,kalmbach-83}),
and then measured ``along'' a different context not containing the observable corresponding to that pure state.
All outcomes except two are discarded~\cite{MR997340,calude:02}, and the two remaining outcomes are mapped onto the symbols  ``0'' and ``1,'' respectively.
If independence of individual ``quantum coin tosses'' is assumed --- a quite nontrivial assumption in view of the Hanbury Brown and Twiss effect and other statistical correlations ---
the concatenation and algorithmic normalization~\cite{von-neumann1,Samuelson-1968} of subsequent recordings of these encoded outcomes
yield an ``absolutely random sequence'' relative to the unprovable axiomatic assumption of quantum randomness.
Since all such operational physical sequences are finite, algorithmic information theory~\cite{calude:02} applies to them in a limited, finite sense.
Particular care should be given to the difficulties in associating an algorithmic information measure to ``nontrivial'' sequences of finite length.

In summary,
there are two principal sources of indeterminism and randomness in physics:
the first source is the deterministic chaos associated with instabilities of classical physical systems,
and with the strong dependence of their future behavior on the initial value;
the second source is quantum indeterminism, which can be subdivided into three subcategories:  random outcomes of individual events,
 complementarity, and
value indefiniteness.

The similarities and differences between classical and quantum randomness can be conceptualized
in terms of two ``black boxes:'' the first one of them --- called the {\em ``Poincar{\'e} box''} ---
containing a classical, deterministic chaotic, source of randomness;
the second  --- called the {\em ``Born box''} ---
containing a quantum source of randomness, such as a quantized system including a beam splitter.
Suppose an agent is being presented with both boxes without any label on, or hint about, them;
i.e., the origin of indeterminism
is unknown to the agent.
In a modified Turing test, the agent's task would be to find out which is the Born and which is
the Poincar{\'e} box by solely observing their output.

It is an open question whether it is possible,
by studying the output behavior of the {\em ``Poincar{\'e} box''} and the {\em ``Born box}''  alone,
to differentiate between them.
In the absence of any criterion, there should not exist any operational  method or procedure
discriminating amongst these boxes.
Both types of indeterminism appear to be based on metaphysical assumptions:
in the classical case it is the existence of continua and the possibility to ``choose''
elements thereof, representing the initial values;
in the quantum case it is the irreducible indeterminism of single events.

It would indeed be tempting also to compare the performance of these physical ``oracles of indeterminism''
with algorithmic cyclic pseydorandom generators, and with irrationals such as $\pi$.
In recent studies~\cite{CDMTCS372} the latter, deterministic, ones seem to be doing pretty well.

In the author's conviction, the postulate of quantum randomness as well as physical randomness emerging from the continuum will be maintained by the community of physicists at large
unless somebody comes up with evidence to the contrary.
This pragmatic interpretation of the phenomena appears reasonable
if and only if researchers are aware of its {\em relativity} with respect to the tests and attempts of falsification involved;
and also acknowledge the tentativeness and conventionality of their assumptions.

{\bf Acknowledgements}
\\
The author gratefully acknowledges the kind hospitality of the {\it Centre for Discrete Mathematics
and Theoretical Computer Science (CDMTCS)} of the {\it Department of Computer Science at
The University of Auckland.}
The {\em ``Poincar{\'e} box''} {\it versus} {\em ``Born box}'' comparison emerged during a walk with Cristian Calude discussing differences of classical {\em versus}
quantum randomness at the Boulevard Saint-Germain in Paris.
This work was also supported by {\it The Department for International Relations}
of the {\em Vienna University of Technology.}


%

\end{document}